\documentclass[preprint,5p]{elsarticle}
\usepackage{stmaryrd}
\usepackage{amsmath}
\usepackage{amssymb}
\usepackage{tabularx}
\usepackage{subfigure}
\usepackage{color}
\usepackage{indentfirst}
\SetSymbolFont{stmry}{bold}{U}{stmry}{m}{n}
\biboptions{sort&compress}
\usepackage{graphicx}
\usepackage{epstopdf}
\usepackage{enumitem}

\numberwithin{equation}{section}

\def\subFS{\scriptscriptstyle{FS}}
\def\subVS{\scriptscriptstyle{VS}}

\newcommand{\mS}{\mathcal{S}}
\newcommand{\bR}{\mathbb{R}}
\newcommand{\rd}{\mathrm{d}}
\newcommand{\dd}[1]{\frac{\rm d}{{\rm d}#1}}
\newcommand{\ddt}{\dd{t}}
\newcommand{\vol}{\operatorname{vol}}
\newcommand{\nn}{\nonumber}

\newcommand{\mF}{\mathcal{F}}
\newcommand{\mG}{\mathcal{G}}
\newcommand{\mA}{\mathcal{A}}
\renewcommand{\vec}[1]{\mathbf{#1}}

\begin{document}

\begin{frontmatter}
\title{Dynamics of Small Solid Particles on Substrates of Arbitrary Topography}
\author[1]{Quan Zhao}
\address[1]{School of Mathematical Sciences, University of Science and Technology of China,  Hefei, China}

\author[2]{Wei Jiang\corref{8}}
\address[2]{School of Mathematics and Statistics,  Wuhan University,  Wuhan, China}
\ead{jiangwei1007@whu.edu.cn}
\cortext[8]{Corresponding author.}

\author[3]{Yan Wang}
\address[3]{ School of Mathematics and Statistics, and Key Lab NAA--MOE Central China Normal University, Wuhan, China}

\author[4]{David J. Srolovitz}
\address[4]{Department of Mechanical Engineering, The University of Hong Kong,  Hong Kong SAR, China}

\author[5]{Tiezheng Qian}
\address[5]{Department of Mathematics, Hong Kong University of Science and Technology, Hong Kong, China}

\author[6]{Weizhu Bao}
\address[6]{Department of Mathematics, National University of Singapore, Singapore}

\begin{abstract}
We study the dynamics of a small solid particle arising from the dewetting of a thin film on a curved substrate driven by capillarity, where mass transport is controlled by surface diffusion.
We consider the case when the size of the deposited particle is much smaller than the local radius of curvature of  the substrate surface.
The application of the Onsager variational principle leads to a reduced-order model for the dynamic behaviour of  particles on arbitrarily curved substrates.
We demonstrate that  particles move toward region of the substrate surface with lower mean curvature with a determined  velocity.
In particular, the velocity is proportional to the substrate curvature gradient and inversely proportional to the size of the particle, with a coefficient that depends on material properties that include the surface energy, surface diffusivity, density, and Young's (wetting) angle.
The reduced model is  validated by comparing with numerical results for the full, sharp-interface model in both two  and three dimensions.

\end{abstract}


\end{frontmatter}

\section{Introduction}

The large and increasing diversity of thin film-based technological applications has led to growing interest in how thin films dewet from or islands forms on substrates (e.g., see \cite{Thompson12, Leroy16,Naffouti17}). Initially continuous thin films are often unstable and  dewet or agglomerate to form isolated particles/island in order to minimize the total interfacial energies.
These observed morphology changes are mainly due to capillary effects, which most commonly occurs via diffusional mass transport along surface \cite{Srolovitz86a}.
In the case of isotropic surface energy, the normal velocity of the evolving interface is proportional to the surface Laplacian of the local mean curvature \cite{Mullins57}.
The dynamics of the contact line, where the thin film/vapor interface meets the substrate, is an additional, important  kinetic feature in the evolution of the morphology.
In particular, at the contact line, the force or line tension balance along the substrate implies an equilibrium contact angle (i.e., Young's law \cite{Young1805}).

A growing body  of experimental and theoretical efforts has focused on the solid state dewetting mechanisms (see e.g., \cite{Srolovitz86a,Ye11b, Amram12,Jiang12,Jiang16,Zucker16,Naffouti17,Jiang19xi,Jiang2018curved,JiangZB20, Boc22stress, Garcke23diffuse}).
In general, the film surface morphology evolution is influenced by many parameters including  film thickness,   film microstructure, surface tension anisotropy, relative surface and interface energies, stress in the film, and the elastic constants of the film and substrate.
More recently, templated dewetting has drawn significant attention; this refers to patterning the shape of the substrate or the film to produce the desired dewetted island microstructures.
Indeed, experiments have shown how topographically patterned substrates can be employed to produce particles of near-uniform size in patterned arrays \cite{Giermann05, Cheng06templated, Wang11, Wang2013solid,Lu16nanostructure, Ruffino17experimental}. Moreover, recent experiments on deposited thin films of platinum on a sinusoidally modulated alumina substrate  have provided the clear evidence about the migration of particles from convex to concave sites of the substrates~\cite{Ahn80}. While most theoretical and simulation studies focus on  flat substrates, relatively little attention has been devoted to dewetting on not-flat substrates (despite their growing interest in experimental studies).
In \cite{Jiang2018curved}, a continuum simulation approach was used to study dewetting on a sinusoidal substrate leading to a period array of islands,  and some related research works can also be found in \cite{Ahn80, Klinger12}. A mathematical understanding of island dynamics on topographically patterned substrates would be valuable  to guide the  precise control of the dewetting process to produce desired self-assembled islands through templating~\cite{Leroy16}.

Recent years have seen a great deal of experimental and theoretical activity focused on the dynamics of liquid droplets on curved substrates, see e.g., \cite{lv14substrate,Galatola18s, Mccarthy19,Chen21self,Sykes22droplet}.
In such a case, the driving force for evolution is related to the local substrate curvature gradient which drives rapid droplet motion.
Chen and Xu \cite{Chen21self} derived a reduced-order model for this type of dynamical system and demonstrated quantitative agreement with experiment.
Their  model was based on the Onsager variational principle \cite{Onsager31a, Onsager31b}, which provides a  useful approximate framework for describing irreversible thermodynamic processes.
This approach is based on the minimization of the Rayleighian, which has contributions from  the free energy change rate and dissipation function which are described in terms of a small number of suitable state variables.
This  leads to a set of ordinary differential equations for these state variables and forms a reduced-order model (see below Section \ref{SEC:Onsager} for a brief description of this  variational principle).

In this paper, we apply the Onsager principle to provide a reduced-order model for the motion  of solid particles on curved substrates, where the dynamics is controlled via surface diffusion (unlike fluid droplets evolving under viscous momentum transport).
For ease of analysis, we  assume that the interface energies are isotropic, the elastic effects of the thin films are negligible and no chemical reactions or phase transformations occur.
We focus on the case of a small particle on a topographically-patterned substrate, in which the length scale of the particle is much smaller than the radius of curvature of any point on the substrate surface.
We show how the total free energy of the system can be related to the substrate curvature  \cite{lv14substrate, Chen21self} and show how the dissipation function is  uniquely determined by the surface normal velocity in two dimensions (2D).
In three dimensions (3D), we develop an alternative approach to compute the dissipation function by connecting it with the Wasserstein metric \cite{Reina15entropy, Van23thermodynamic}, which leads to a constrained minimization problem.
We demonstrate that in both 2D and 3D that the new derived model is both simple and quantitative  via several examples.

The remainder of the paper is organized as follows.
In Section~\ref{SEC:fmodel}, we introduce the full sharp-interface model for the dewetting of solid thin films on a curved substrate and include a brief review of the Onsager variational principle.
Next, in Section \ref{SEC:2D}, we apply this principal to derive a reduced-order model for the dynamics of a 2D particle and validate the model by the numerical results from solving the full model.
In Section \ref{SEC:3D}, we generalize this approach to three dimensions.  In Section \ref{SEC:GEN}, we provide some generalizations of the proposed approach to the cases
of chemically inhomogeneous flat substrates and anisotropic solid particles.
Finally, we draw some conclusions in Section \ref{SEC:CON}.

\section{Preminary}

\label{SEC:fmodel}

In this section, we first introduce the full sharp-interface model for the dewetting system and then give a short review of the Onsager principle.
\subsection{The full sharp-interface model}
\label{SEC:fsharpmodel}
\begin{figure}[!htp]
\centering
\includegraphics[width=.45\textwidth]{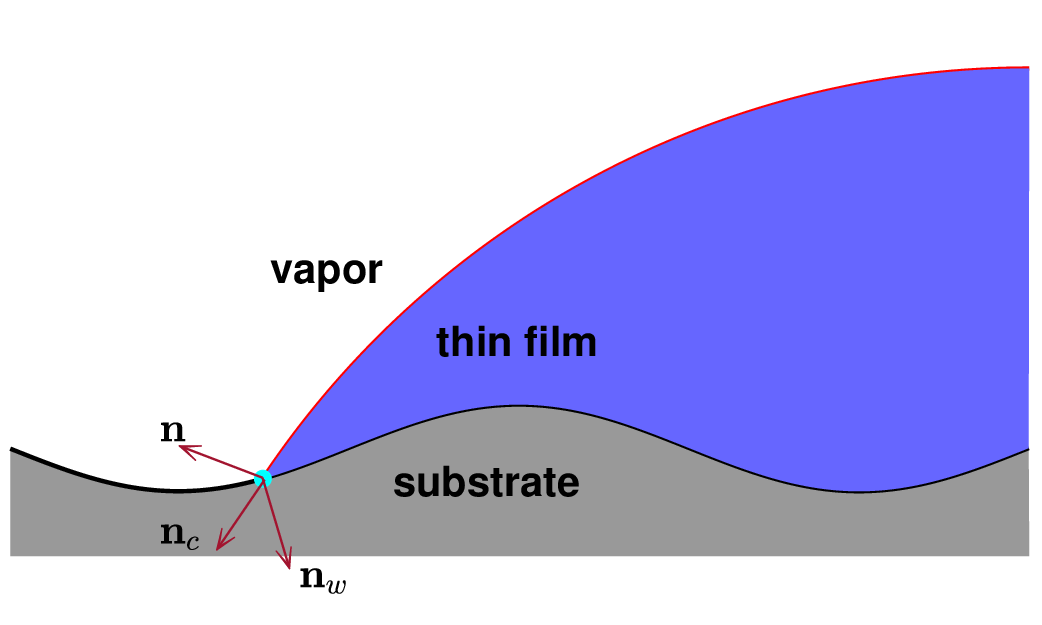}
\caption{A schematic illustration of a solid particle on a curved substrate.}
\label{fig:mod}
\end{figure}

Consier a solid thin film deposited on a curved substrate, as shown in Fig.~\ref{fig:mod}.
The film/vapor interface is represented by an open hypersurface $\mS(t)$ in $d=2$ or $3$ dimensions,  $d\in\{2,3\}$.
We assume  a parameterization of the film surface $\mS(t)$ on the reference domain $\mathcal{O}\subset\bR^{d-1}$ (1 dimension lower than the space)  given by
\begin{equation}
\vec X(\vec\rho, t): \mathcal{O}\times[0,T]\mapsto\bR^d.\nn
\end{equation}
The velocity of the interface $\mS(t)$  is
\begin{equation}
\vec V(\vec X(\vec \rho,t), t) = \partial_t\vec X(\vec\rho,t)\quad\mbox{for}\quad\vec X\in\mS(t).\nn
\end{equation}
Let $\vec n$ and $\mathcal{H}$ be the unit normal and mean curvature of the interface $\mS(t)$, respectively.
Assuming isotropic diffusion and surface tension, the interface dynamics can be expressed as \cite{Mullins57,Cahn94}
\begin{subequations}\label{eqn:fmod}
\begin{align}\label{eq:fmod1}
V_n &=  -\Omega_0\nabla_s\cdot\vec j,\\
\label{eq:fmod2}
\vec j &= -\frac{D_s\nu}{k_b\,T}\nabla_s\mu,\\
\mu &= \Omega_0\gamma_0\mathcal{H},
\label{eq:fmod3}
\end{align}
\end{subequations}
where $V_n=\vec V\cdot\vec n$ is the velocity of the  interface in the direction of $\vec n$, $\vec j$ is the flux of the surface atoms, $\mu$ is the chemical potential of a film atom on the surface and $\nabla_s$ is the surface gradient operator.
The physical parameters/constants are the volume per atom of the film material $\Omega_0$ , $D_s$ is the surface diffusivity, $k_b T$ is the thermal energy,
$\nu$ is the number of diffusing atoms per unit area (in the direction normal to the surface flux vector), and $\gamma_0$ is the isotropic surface energy density.
Combining terms,  \eqref{eqn:fmod} can be rewritten as
\begin{equation}
V_n = B\gamma_0\,\Delta_s\mathcal{H},\nn
\end{equation}
where $B=\frac{D_s\nu\Omega_0^2}{k_b\,T}$ is a material constant and $\Delta_s$ is the Laplace-Beltrami (Laplacian) operator.

At the contact line  $\Gamma(t)$ where the film/varpor interface meets the substrate, we impose the following boundary conditions:
\begin{subequations}
\begin{itemize}
\item [(i)]attachment condition
\begin{equation}
\vec V\cdot\vec n_w = 0; \label{eq:bd1}
\end{equation}
\item [(ii)]contact angle condition
\begin{equation}
\vec n \cdot\vec n_w + \cos\theta_i=0;\label{eq:bd2}
\end{equation}
\item [(iii)]zero-flux condition
\begin{equation}
\vec n_c\cdot\vec j = 0.\label{eq:bd3}
\end{equation}
\end{itemize}
\end{subequations}
Here, $\vec n_w$ is the substrate unit normal (positive pointing towards the substrate interior) and $\vec n_c$ is the conormal vector of $\Gamma(t)$, as shown in Fig.~\ref{fig:mod}.
Moreover, $\theta_i$ is the equilibrium, isotropic Young angle, which satisfies
\begin{equation}
\cos\theta_i=(\gamma_{_{\subVS}}-\gamma_{_{\subFS}})/ \gamma_0,\nn
\end{equation}
with $\gamma_{_{\subVS}}$ and $\gamma_{_{\subFS}}$ representing the varpor/substrate and film/substrate surface energy densities, respectively. Condition (ii) can be interpreted as the contact angle condition; this  leads to the Young angle  $\theta_i$ between $\vec n$ and -$\vec n_w$ at the contact line.
We note that under some conditions, the dynamic contact angle may differ from the Young's  contact angle condition (e.g., see \cite{Jiang2018curved, Karim22}).

The total free energy of the dynamic system is given by
\begin{equation}
W(t) = \gamma_0|\mS(t)|-\gamma_0\cos\theta_i A_{\rm sub}(\Gamma(t)),\label{eq:fenergy}
\end{equation}
where $|\mS(t)|$ is the surface area of $\mS(t)$, and $A_{\rm  sub}(\Gamma(t))$ represents the substrate surface area covered by the film/island (i.e., enclosed by $\Gamma(t)$).
The free energy of the evolving film/substrate system satisfies
\begin{equation}\label{eq:dtw}
\ddt W(t) = -\frac{k_b\,T}{D_s\,\nu}\int_{\mS(t)}|\vec j|^2\,\rd S\leq 0.
\end{equation}
We also assume that the volume of the film material is conserved (we do not consider phase transformations or strains); i.e., $\vol(\Omega(t)) = \vol(\Omega(0))$ for all time $t\geq0$.

\begin{figure*}[!htp]
\centering
\includegraphics[width=.40\textwidth]{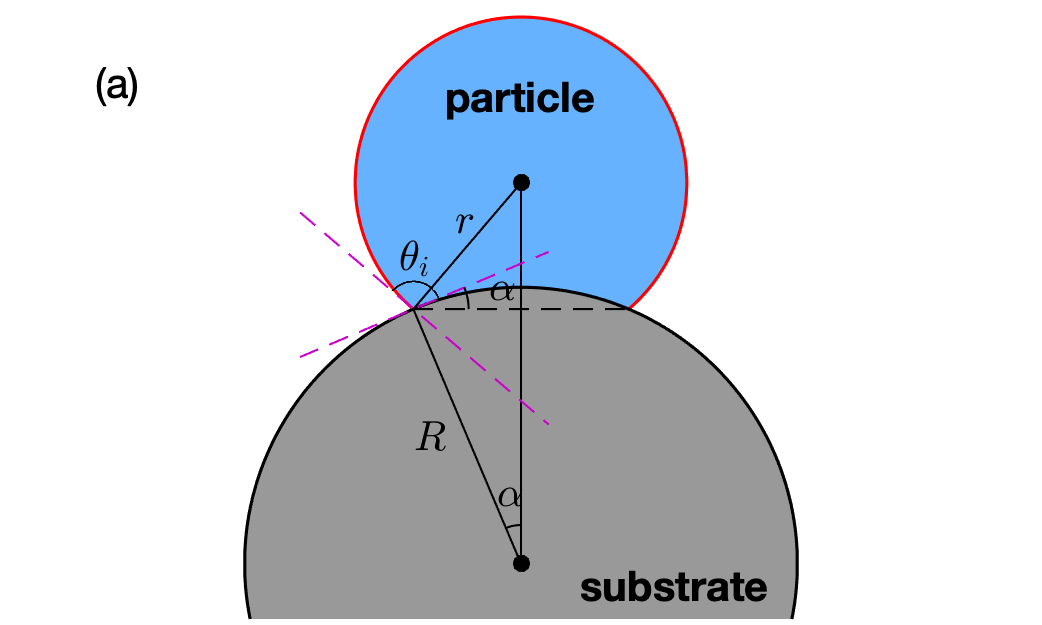}
\includegraphics[width=.40\textwidth]{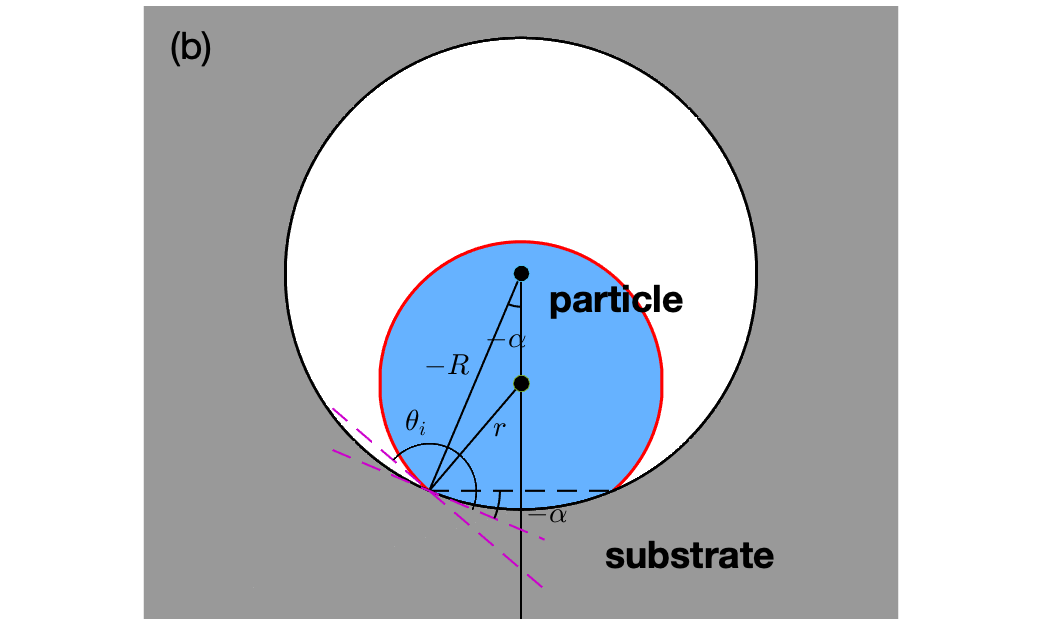}
\caption{Local approximation of the substrate/particle for a substrate surface of (a) positive and (b) negative curvature. Here $\theta_i$ refers to Young contact angle, and $R\sin\alpha>0$ with $|\alpha|\ll 1$.  }
\label{fig:1}
\end{figure*}

\subsection{Onsager variational principle}\label{SEC:Onsager}
The Onsager variational principle was first formulated~\cite{Onsager31a,Onsager31b} based on the reciprocal symmetry in a linear irreversible thermodynamic process.
This fundamental principle provides a general framework to describe non-equilibrium kinetics in cases where linear response  is applicable and has found  wide applications in fluid dynamics~\cite{Qian06,Qian17,Xu16,Di18,Man16, Zhang22effective},  soft matter physics~\cite{Doi11,Doi13book,Doi15} and solid-state deweting \cite{JZOnsager}. We first provide a short review of this principle.

Consider an isothermal system described by a set of time-dependent state variables
\[\beta(t) = (\beta_1(t),\beta_2(t),\ldots, \beta_n(t)),\]
and  $\dot{\beta}(t)=(\dot{\beta}_1(t),\dot{\beta}_2(t),\ldots,\dot{\beta}_n(t))$ are the rates of change of these state variables (the raised dot ``$\cdot$'' denotes a time derivative).
We further introduce $W(\beta)$ as the total free energy of the system, then
the rates $\{\dot{\beta}_i\}$ are  determined by miniminzing the Rayleighian \cite{Doi15,Xu16,Suo97}
\begin{equation}
\mathcal{R}(\dot{\beta},\beta)=\dot{W}(\beta, \dot{\beta})+\Phi(\dot{\beta},\dot{\beta}),
\label{eqn:rayleigh}
\end{equation}
where $\dot{W}(\beta, \dot{\beta})=\sum_{i=1}^n(\partial W / \partial \beta_i)\dot{\beta_i}$ is the rate of change of the total free energy $W$
and $\Phi(\dot{\beta},\dot{\beta})$ is the dissipation function.

In the linear response regime, the dissipation function is a quadratic function of the rates $\{\dot{\beta}_i\}$
\begin{equation}
\Phi(\dot{\beta},\dot{\beta})=\frac{1}{2}\sum_{i=1}^n\sum_{j=1}^n\lambda_{ij}(\beta)\dot{\beta}_i\dot{\beta}_j,
\label{eq:dissf}
\end{equation}
where the friction coefficients $\{\lambda_{ij}\}$ form a positive definite, symmetric  matrix.
Minimizing the Rayleighian  \eqref{eqn:rayleigh} with respect to  rates $\{\dot{\beta}_i\}$  yields the kinetic equations
\begin{equation}
-\frac{\partial W}{\partial \beta_i}=\sum_{j=1}^n\lambda_{ij}\dot{\beta}_j,\qquad i=1,2,\ldots,n,
\label{eq:kineticEq}
\end{equation}
which precisely gives the force balance between the reversible force $-\partial W/ \partial \beta_i$ and the dissipative force $-\partial \Phi / \partial \dot{\beta}_i$. Multiplying \eqref{eq:kineticEq} by $\beta_i$ and summing (and  recalling \eqref{eq:dissf})  yields
\begin{equation}
\Phi(\dot{\beta},\dot{\beta})=-\frac{1}{2}\dot{W}(\beta,\dot{\beta}).
\label{eq:dispf}
 \end{equation}
This means that the dissipation function is half the rate of the free energy dissipation.
Physically, the variational principle  for {\it isothermal} systems can  also be derived from the maximization of the Onsager-Machlup action that is used for more general {\it non-isothermal} systems~\cite{Onsager31a,Onsager31b}.

In the present work, we apply the Onsager variational principle to describe the dynamics
of a small particle migrating on a curved substrate.
We  focus on  cases in which the full model, introduced in \S\ref{SEC:fsharpmodel}, can be approximately described by a finite set of suitable state variables.
Application of the Onsager principle  enables us to obtain a reduced model for a continuous dissipative system which is governed by a set of ordinary differential equations for a few state variables.
Note that the full model satisfies the energy dissipation law in \eqref{eq:dtw} and  can be obtained as well by applying the Onsager principle in the linear response regime,
where the constitutive equation \eqref{eq:fmod2} is derived for the flux $\vec j$.

\section{Dynamics in two dimensions}\label{SEC:2D}

We now apply the Onsager variational principle in \S\ref{SEC:Onsager} to derive a reduced-order model for the dynamics of a particle on a substrate in 2D.
The reduced model is then numerically validated by comparisons  with the full model in \S\ref{SEC:fsharpmodel}.

\subsection{A reduced-order model}

We assume that the particle/island  is much smaller than the radius of curvature  of the substrate.
The separation of these two length scales leads to two distinct time scales:
one for the island to establish the circular shape and the other for the island to migrate along the substrate.
In particular, the latter is much slower than the former.
Therefore, it is reasonable to assume that the particle will remain circular during its (relatively) slow motion along the substrate.

As shown in Fig.~\ref{fig:1}(a), we assume that the film/vapor interface of the particle is given by section of a  (small) circle of radius $r(t)$ and the substrate is {\it locally} approximated by a (large) circle of radius $R(P)$, where $P(t)$ represents the intersection point of the substrate and the straight line that connects the centers of the two circles (at all times $R\gg r$).
This gives rise to a parameterization of the interface profile $\vec X(\theta, t)=(x(\theta, t), y(\theta, t))^T$ as
\begin{equation}
\left\{
\begin{array}{l}
x(\theta,t)=P(t)+r(t)\sin\theta,\\[0.8em]
y(\theta,t)=r(t)(\cos\theta-\cos\hat{\theta}),
\end{array}
\right.\quad \theta\in[-\hat{\theta},\hat{\theta}],
\label{eqn:profile}
\end{equation}
where $\hat{\theta}=\theta_i+\alpha(P)$ with  $r\sin\hat{\theta} = R\sin\alpha$.
The conserved particle area is
\begin{equation}
A_0= r^2\zeta(\hat{\theta}) - R^2\zeta(\alpha),
\label{eq:2dvol}
\end{equation}
where $\zeta(\alpha) = \alpha -\cos \alpha\sin \alpha$.
For $0 \leq \alpha \ll 1$, this yields the following  identities
\begin{subequations}\label{eqn:RrA}
\begin{align}
\frac{\sqrt{A_0}}{R}&:= \sqrt{\frac{\sin^2\alpha}{\sin^2\hat\theta}\,\zeta(\hat{\theta}) - \zeta(\alpha)} ,\\
\frac{r}{\sqrt{A_0}}&:=\sqrt{\frac{\sin^2\alpha}{\sin^2\alpha\,\zeta(\hat{\theta}) - \sin^2\hat{\theta}\,\zeta(\alpha)}}.
\end{align}
\end{subequations}
Taylor expanding \eqref{eqn:RrA} about $\alpha = 0$ gives
\begin{subequations}\label{eq:RRExpan}
\begin{align}
\label{eq:RA0}
\frac{\sqrt{A_0}}{R}&= \frac{\sqrt{\zeta(\theta_i)}}{\sin\theta_i}\alpha + O(\alpha^2), \\
\frac{r}{\sqrt{A_0}} &= \frac{1}{\sqrt{\zeta(\theta_i)}} + O(\alpha).
\label{eq:rA0}
\end{align}
\end{subequations}

Recalling \eqref{eq:fenergy}, the total free energy of the approximate 2D  system can be written as
\begin{equation}
W = 2\gamma_0(r\hat{\theta} - R\cos\theta_i\,\alpha).\label{eq:2denergy}
\end{equation}
Using \eqref{eq:RRExpan} in \eqref{eq:2denergy}  yields
\begin{equation}
W(P)=2\gamma_0\sqrt{A_0}\,\left(\sqrt{\zeta(\theta_i)} + \frac{\sin^2\theta_i}{3\,\sqrt{\zeta(\theta_i)}}\alpha + O(\alpha^2)\right).\nn
\end{equation}
Taking the time derivative of the total free energy and using \eqref{eq:RA0}, we obtain
\begin{align}
\dot{W}(P, \dot{P}) &=  \frac{2\gamma_0\,\sqrt{A_0}\,\sin^2\theta_i}{3\,\sqrt{\zeta(\theta_i)}} \alpha^\prime(P)\dot{P} +O(\alpha^2)\nn\\
&\approx \frac{2\gamma_0\,A_0\,\sin^3\theta_i\,}{3\,\zeta(\theta_i)}\kappa^\prime(P)\,\dot{P},\label{eq:finalWdot}
\end{align}
where we introduced the substrate curvature (in 2D) $\kappa(P) = R(P)^{-1}$  and   primes  denote  derivatives with respect to $P$.

To this point, we  restricted ourselves to the case when the curvature of the substrate is positive at the point of interest, implying that $R(P(t)) > 0$ and $\alpha\geq 0$, see Fig.~\ref{fig:1}(a).
When the substrate curvature is negative, $R(P(t)) < 0$;  equations \eqref{eqn:profile}, \eqref{eq:2dvol}, \eqref{eq:2denergy} and \eqref{eq:finalWdot} hold as well except that  $0<-\alpha\ll 1$, as shown in Fig.~\ref{fig:1}(b).

Recalling \eqref{eq:fenergy}, we write the dissipation function  \eqref{eq:dispf} as
\begin{equation}
\label{eq:Phi2d}
\Phi=\frac{1}{2}\frac{k_b\,T}{D_s\,\nu}\int_{\mS(t)}|\vec j|^2\,\rd S =\frac{1}{2}\frac{k_b\,T}{D_s\,\nu}\int_{\mS(t)}|J|^2\,\rd S,
\end{equation}
where $J(\theta)=\vec j\cdot\boldsymbol{\tau}$ is the magnitude of the flux and $\boldsymbol{\tau}$ is the unit interface tangent.
Using the kinematic equation \eqref{eq:fmod1}, we find
\begin{equation}
\partial_s J(\theta) = -\frac{1}{\Omega_0}V_n(\theta)\quad\mbox{with}\quad J(\pm\hat{\theta})=0,
\label{eq:flux}
\end{equation}
where $\theta\in[-\hat{\theta},\hat{\theta}]$ and $V_n(\theta)$ is
\begin{equation}
V_n(\theta) = \dot{P} \sin\theta + \dot{r}(1-\cos\theta_i\cos\theta),\label{eq:vn}
\end{equation}
on recalling \eqref{eqn:profile} and $\vec n = (\sin\theta, \cos\theta)^T$.
Using \eqref{eq:vn} and integrating \eqref{eq:flux} then yields
\begin{align}
J(\theta) & = -\frac{1}{\Omega_0}\int_{-\hat{\theta}}^\theta r V_n(\theta)\,\rd\theta \nn\\
&= -\frac{1}{\Omega_0}\dot{P} r(\cos\theta_i - \cos\theta)+ O(\alpha), \label{eq:fluxne}
\end{align}
where we note $\dot{r} = O(\alpha)$ because of \eqref{eq:rA0}.
Inserting \eqref{eq:fluxne} into \eqref{eq:Phi2d} and using \eqref{eq:rA0},  we find
\begin{align}
\Phi(\dot{P}) &= \frac{1}{2}\frac{k_b\,T}{D_s\,\nu}\int_{-\hat{\theta}}^{\hat{\theta}}|J^2|\,r\rd\theta\nn\\
&= C_{2\rd}^0(\theta_i)\,B^{-1}\,r^3\dot{P}^2 +O(\alpha^2)\nn\\
&\approx C_{2\rd}^0(\theta_i)\,B^{-1}\sqrt{\frac{A_0^3}{\zeta^3(\theta_i)}}\dot{P}^2,\label{eq:finalPhi}
\end{align}
where $B={D_s\nu\Omega_0^2}/{k_B\,T}$ and \[C_{2\rd}^0(\theta_i)=\frac{1}{2}(\theta_i + 2\theta_i\cos^2\theta_i - 3\sin\theta_i\cos\theta_i).\]
Note that for the dynamics in 2D, the flux $J(\theta)$ is completely determined from
$\dot{P}$ and $V_n(\theta)$, and hence the coefficient of $\Phi(\dot{P})\propto\dot{P}^2$ is
readily obtained from \eqref{eq:finalPhi}
(this is not the case in 3D).

Using \eqref{eq:finalWdot} and \eqref{eq:finalPhi} in \eqref{eqn:rayleigh}, and applying the Onsager principle by minimizing the Rayleighian $\mathcal{R}$ with respect to $\dot{P}$, we thus obtain the following ODE for the state variable $P$~(see also Equation (16) in \cite{Jiang2018curved}):
\begin{equation}\label{eq:2ddynamic}
\begin{cases}
    \displaystyle \frac {{\rm d} P}{{\rm d} t} = - \frac{B\gamma_0\,C_{2\rd}(\theta_i)}{\sqrt{A_0}}\,\kappa\,' (P),\\
    P(0) = P_0,
\end{cases}
\end{equation}
where
\begin{equation}
C_{2\rd}(\theta_i) =\frac{2\sin^3(\theta_i)\,\sqrt{\theta_i- \cos\theta_i\sin\theta_i}}{3(\theta_i+2\theta_i\cos^2\theta_i - 3\sin\theta_i\cos\theta_i)}.
\end{equation}

Equation \eqref{eq:2ddynamic} prescribes the velocity of the particle on the substrate  when $|\alpha|\ll 1$.
It shows that the velocity is proportional to the substrate curvature gradient and is inversely proportional to the length scale of the particle, i.e., $\sqrt{A_0}$.
Moreover, the coefficient $C_{2\rd}(\theta_i)$ is a monotone decreasing function for $\theta_i\in[0,\pi]$.
This implies that the velocity tends to  zero as $\theta_i$  approaches $\pi$. Physically, a larger contact angle $\theta_i$ implies a weaker coupling between the particle and the substrate (with the contact length or area approaching zero for $\theta_i\to\pi$), and hence the migration velocity $\dot{P}$ becomes less responsive to substrate curvature gradients.

\subsection{Numerical  validation}\label{SEC:NUM}

We choose the length scale $L_0$, time scale ${L_0^4}/{B\gamma_0}$ and use the quantities with hats ($\hat{\cdot}$) to denote dimensionless physical quantities.
The full model \eqref{eqn:fmod} can then be rewritten in  dimensionless form  as
\begin{equation}\label{eqn:dfmod}
    \hat{V}_n = \hat{\Delta}_s\hat{\mathcal{H}},
\end{equation}
with boundary conditions at the contact line:
\begin{subequations}
\begin{itemize}
\item [(i)]attachment condition
\begin{equation}
\hat{\vec V}\cdot\vec n_w = 0; \label{eq:dbd1}
\end{equation}
\item [(ii)]contact angle condition
\begin{equation}
\vec n \cdot\vec n_w + \cos\theta_i=0;\label{eq:dbd2}
\end{equation}
\item [(iii)]zero-flux condition
\begin{equation}
\vec n_c\cdot\hat{\nabla}_s\hat{\mathcal{H}} = 0.\label{eq:dbd3}
\end{equation}
\end{itemize}
\end{subequations}
The reduced model, in  dimensionless form, is
\begin{equation}\label{eq:d2ddynamic}
\begin{cases}
    \displaystyle \frac {{\rm d} \hat{P}}{{\rm d} \hat{t}} = - \frac{\,C_{2\rd}(\theta_i)}{\sqrt{\hat{A}_0}}\,\hat{\kappa}\,' (\hat{P}),\\
    \hat{P}(0) = \hat{P}_0.
\end{cases}
\end{equation}
We then conduct numerical comparisons between the two models.
We employ a parametric finite element method to solve the full sharp-interface model (see \cite{Jiang2018curved, Barrett20}).
The system of ODEs for the reduced model are solved via the forward Euler method.

\begin{figure}[!htp]
\centering
 \includegraphics[width = .475\textwidth]{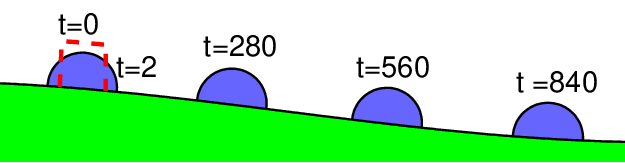} \\[0.5em]
\includegraphics[width=.225\textwidth]{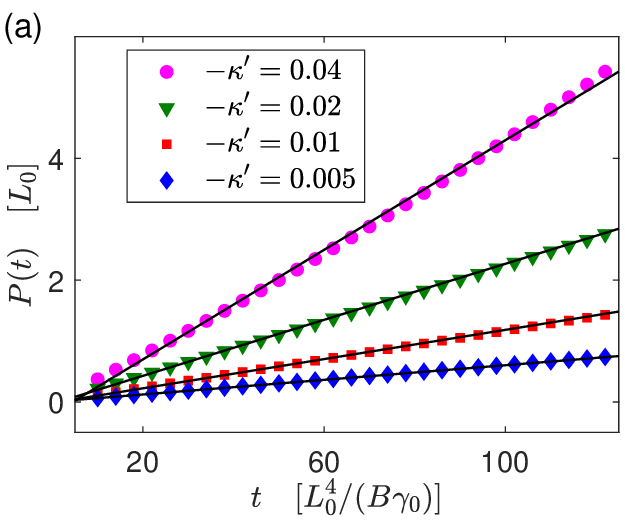}
\hspace{0.05cm}
\includegraphics[width=.225\textwidth]{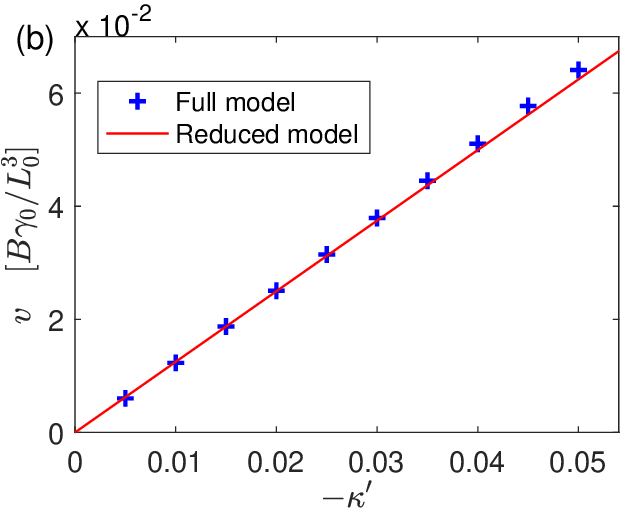}
\includegraphics[width=.225\textwidth]{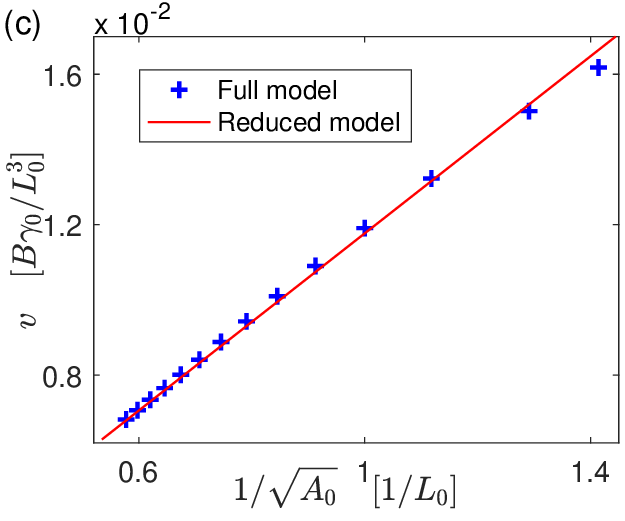}\hspace{0.05cm}
\includegraphics[width=.225\textwidth]{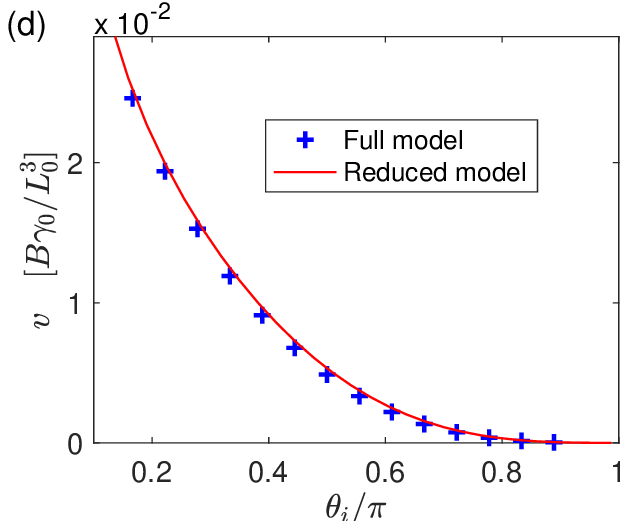}
\caption{On the top panel shows a full model simulation of  the migration of a ``small'' ($\hat{A}_0=0.4$) solid particle on a curved substrate ($\hat{k}^\prime(s)=-0.01$), where $\theta_i = \pi/2$, and the red dashed line represents the shape/position of the initial particle. (a): The position of the particle against time for substrates of different curvature gradients with $\hat{A}_0=1$, $\theta_i=\pi/3$, where the solid lines are linear fits to the displacement. (b): The velocity of the particle $v = |{dP(t)}/{dt}|$ versus the substrate curvature gradient (${A}_0=1$, $\theta_i=\pi/3$). The symbols represent full model simulations and the red line is an analytical solution of the reduced model. (c): The velocity of the particle versus  particle area ($\hat{\kappa}^\prime=0.01$, $\theta_i=\pi/3$). (d): The velocity of the particle versus $\theta_i/\pi$  ($\hat{\kappa}^\prime=-0.01$, $\hat{A}_0=1$). }
\label{fig:3}
\end{figure}

We first perform a full model simulation of the migration of a small particle ($\hat{A}_0=0.4$) on a substrate, which is modeled by a curve satisfying $\hat{\kappa}^\prime(s)=-0.01$, where  $s$ is the arc length parameter.
Initially, we place a square particle at a position with $\hat{\kappa}=0.05$ and then capture the dynamics of the particle  to a position with $\hat{\kappa} = -0.05$.
As shown in Fig.~\ref{fig:3}, we see observe that the particle quickly adopts a circular shape and maintains it while gradually migrates towards a position of lower curvature.

We next conduct a series of simulations for a particle on substrates with different curvature gradients, where the other parameters are fixed ($\hat{A}_0=1$, $\theta_i=\pi/3$).
The numerical results obtained from the full model are presented in Fig.~\ref{fig:3}(a).
We observe that the position of the particle varies approximately linearly with time and the speed of the particle is proportional to the substrate curvature gradient.
In particular, the speed of the particle is very similar to the analytical results from the reduced model, as shown in Fig.~\ref{fig:3}(b).

To further validate the reduced model, we next study the dependence of the particle velocity on $A_0$ and $\theta_i$, and the curvature gradient of the substrate is fixed as $\hat{\kappa}^\prime = -0.01$. It can be readily seen that the velocity of the particle is inversely proportional to $\sqrt{A_0}$ (Fig.~\ref{fig:3}(c)) and proportional to $C_{2\rd}(\theta_i)$ (Fig.~\ref{fig:3}(d)). In particular, the results show excellent quantitative agreement with the reduced model.

\begin{figure}
\centering
\includegraphics[width=.45\textwidth]{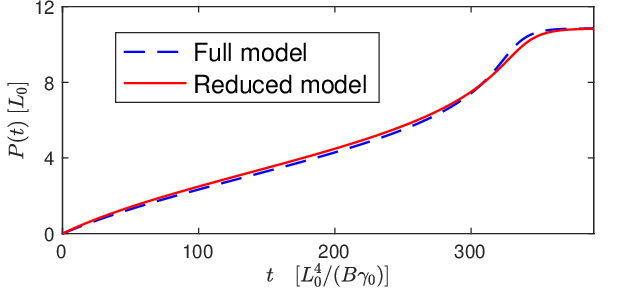}
\caption{The evolution  of the particle position  on a sinusoidal substrate $y=4\sin(x/4)$ with  $\hat{A}_0=1$ and $\theta_i=\pi/3$.
The zero of time $\hat{t}=0$ was set to that at which the square island becomes nearly circular (i.e., at $\hat{t}=50$ in the full model evolution).}
\label{fig:4}
\end{figure}

Finally, we investigate the dynamics of a particle on a general sinusoidal substrate described by $y = 4\sin(x/4)$ with  $\hat{A}_0=1$ and $\theta_i=\pi/3$.
The numerical results from the full sharp-interface model are compared with those from solving the system of ODEs in the reduced model, \eqref{eq:2ddynamic}.
The agreement between the full and reduced model results are in excellent quantitative agreement as observed in Fig.~\ref{fig:4}.
This not only validates our reduced-order model in \S\ref{SEC:2D}, but also
verifies the accuracy of the numerical results from the full sharp-interface model. Moreover, we note that these theoretical results are also consistent with those reported in \cite{Ahn80, Klinger12}.

\section{Dynamics in three dimensions}\label{SEC:3D}

\subsection{The reduced model}
\label{SEC:3dred}

Inspired by the work in \S\ref{SEC:2D}, we now consider the extension of the reduced model from 2D to 3D.
Given the position $\vec P =(P_x, P_y, 0)^T$ on the substrate surface, we  locally approximate the substrate at $\vec P$  as a sphere of radius $R(\vec P)=\frac{1}{\kappa(\vec P)}$, where $\kappa(\vec P)=\frac{1}{2}(\kappa_1 + \kappa_2)$ is the mean curvature of the substrate with $\kappa_1$ and $\kappa_2$ being the two principal curvatures. We parameterize the interface $\mS(t)$ by $\vec X:=\vec X(\theta,\phi,t)$
\begin{equation}
\left\{
\begin{array}{l}
x(\theta,\varphi,t)=P_x(t)+r(t)\sin\theta\cos\phi,\\[0.8em]
y(\theta,\varphi,t)=P_y(t)+r(t)\sin\theta\sin\phi,\\[0.8em]
z(\theta,\varphi, t)=r(t)(\cos\theta-\cos\hat{\theta}),
\end{array}
\right.
\label{eqn:3dprofile}
\end{equation}
for $\theta\in[0,\hat{\theta}]$ and $\phi\in[0,2\pi]$, where $\hat\theta = \theta_i + \alpha$ with $\alpha$ satisfying $R\sin\alpha = r\sin\hat\theta$.
The volume of the particle is the volume difference of two spherical caps
\begin{equation}\label{eq:3dvol}
V_0 =\frac{\pi}{3}\left(r^3\,\eta(\hat\theta)-R^3\,\eta(\alpha)\right),
\end{equation}
with $\eta(\alpha) = (1-\cos \alpha)^2(2+\cos \alpha)$.
The total free energy  of the 3D system is
\begin{equation}\label{eq:3denergy}
W = 2\pi\gamma_0[r^2\,(1-\cos\hat\theta) - R^2\,\cos\theta_i\,(1-\cos\alpha)].
\end{equation}
Using \eqref{eq:3dvol} in \eqref{eq:3denergy} leads to
\begin{align}
W(\vec P) =& 2\gamma_0\sqrt[3]{\frac{9\,V_0^2\,\pi}{\eta^2(\theta_i)}}\,(1-\cos\theta_i-\frac{1}{2}\cos\theta_i\sin^2\theta_i)\nn\\ &+\frac{3\gamma_0 V_0}{2}\frac{(1+\cos\theta_i)^2}{(2+\cos\theta_i)}\frac{1}{R(\vec P)} + O(\alpha^2).\label{eq:3denergyapp}
\end{align}
Using the fact that the substrate curvature $\kappa(\vec P) = {1}/{R(\vec P)}$, the time derivative of the energy becomes
\begin{equation}
\dot{W}(\vec P, \dot{\vec P}) \approx \frac{3\gamma_0 V_0}{2}\frac{(1+\cos\theta_i)^2}{(2+\cos\theta_i)}\nabla_\Gamma\kappa\cdot\dot{\vec P},\label{eq:3dWdot}
\end{equation}
where $\nabla_\Gamma$ represents the  curvature gradient along the substrate surface.

In contrast to the 2D case,  the flux contribution on the spherical cap in  3D cannot be uniquely determined by Eq.~\eqref{eq:fmod1} although the normal velocity of the interface is given.
Nevertheless, the dissipation function can be connected with and interpreted as the Wasserstein distance in the framework of minimum dissipation \cite{Van23thermodynamic}.
This leads to a constrained minimization problem:
\begin{subequations}\label{eqn:cmin}
\begin{alignat}{2}
&\min_{\vec j}\Phi(\dot{\vec P})=\frac{1}{2}\frac{k_b\,T}{D_s\,\nu}\int_{\mS(t)}|\vec j|^2\,\rd S &&\\
&\mbox{with}\quad\nabla_s\cdot\vec j = -\frac{\dot{\vec P}\cdot\vec n}{\Omega_0}\quad &&\mbox{on}\quad\mS(t),\\
&\mbox{and}\quad \vec j\cdot\vec n_c=0\quad &&\mbox{at}\quad\Gamma(t).
\end{alignat}
\end{subequations}
in which the dissipation function $\Phi(\dot{\vec P})$ is obtained through minimization
with respect to ${\vec j}$ subject to the constraint imposed by the continuity equation.
This system satisfies  rotational invariance, meaning that
\[\Phi(\dot{\vec P})\propto |\dot{\vec P}|^2.\]
We then introduce the dimensionless flux
\begin{equation}\label{eq:dflux}
    \tilde{\vec j}=\frac{\vec j}{r\,\Omega^{-1}\,|\dot{\vec P}|},
\end{equation}
to obtain
\begin{align}
\Phi(\dot{\vec P})&= \frac{1}{2}\frac{k_b\,T}{D_s\,\nu}\int_{\mS(t)}r^2\Omega^{-2}|\dot{\vec P}|^2\,|\tilde{\vec j}|^2\,\rd S\nn\\
&=\frac{1}{2}\frac{k_b\,T}{D_s\nu\Omega_0^2} r^4\,|\dot{\vec P}|^2 \int_{\tilde{\mS}(t)}|\tilde{\vec j}|^2\,\rd \tilde{\mS}\nn\\
&=\frac{k_b\,T}{D_s\nu\Omega_0^2}\,\left(\frac{3\,V_0}{\pi\,\eta(\theta_i)}\right)^\frac{4}{3}\,|\dot{\vec P}|^2\,m(\theta_i),
\label{eq:3dPhidot}
\end{align}
where  $\tilde{\mS}(t)$ is  $S(t)$ scaled by the dimension $r$ and $m(\theta_i)=\frac{1}{2}\int_{\tilde{\mS}(t)}|\tilde{\vec j}|^2\,\rd \tilde{\mS}$ is a dimensionless function of the Young angle $\theta_i$.
In practice, we compute $m(\theta_i)$ via the  minimization problem \eqref{eqn:cmin} as described in \ref{app:cmin}.

Combining Eqs.~\eqref{eq:3dWdot} and \eqref{eq:3dPhidot} with Eq.~\eqref{eqn:rayleigh} and applying the Onsager principle (minimizing the Rayleighian $\mathcal{R}$ with respect to $\dot{\vec P}$), we  obtain our reduced-order model for the motion of a particle in 3D.
The particle velocity is
\begin{equation}\label{eq:3ddynamic}
\begin{cases}
    \displaystyle \frac {{\rm d} \vec P}{{\rm d} t} = - \frac{B\gamma_0\,C_{3\rd}(\theta_i)}{\sqrt[3]{V_0}}\,\nabla_\Gamma\kappa(\vec P),\\
    \vec P(0) = \vec P_0,
\end{cases}
\end{equation}
where
\begin{equation}
C_{3\rd}(\theta_i) = \frac{\pi}{4}\frac{(1-\cos^2\theta_i)^2}{m(\theta_i)}\sqrt[3]{\frac{\pi\,\eta(\theta_i)}{3}}.
\end{equation}
Note that the application of the Onsager variational principle to the dynamics in 3D
consists of {\it two steps}:
(i) Obtaining the dissipation function $\Phi(\dot{\vec P})$  by minimizing the rate of dissipation
with respect to $\vec j$ subject to the constraint imposed by the continuity equation
for a prescribed $\dot{\vec P}$;
(ii) Determination of the migration velocity $\dot{\vec P}$ by minimizing the Rayleighian $\mathcal{R}$ with respect to $\dot{\vec P}$.

The particle moves in a direction opposite the curvature gradient, meaning that the trajectory of the particle is purely determined by the substrate topography.
While the other physical parameters (e.g.,$B$, $\gamma_0$, $V_0$ and $\theta_i$) only affect how fast the particle moves in this particular trajectory.

\subsection{Numerical validation}
\begin{figure*}[!htp]
\centering
\includegraphics[width=0.9\textwidth]{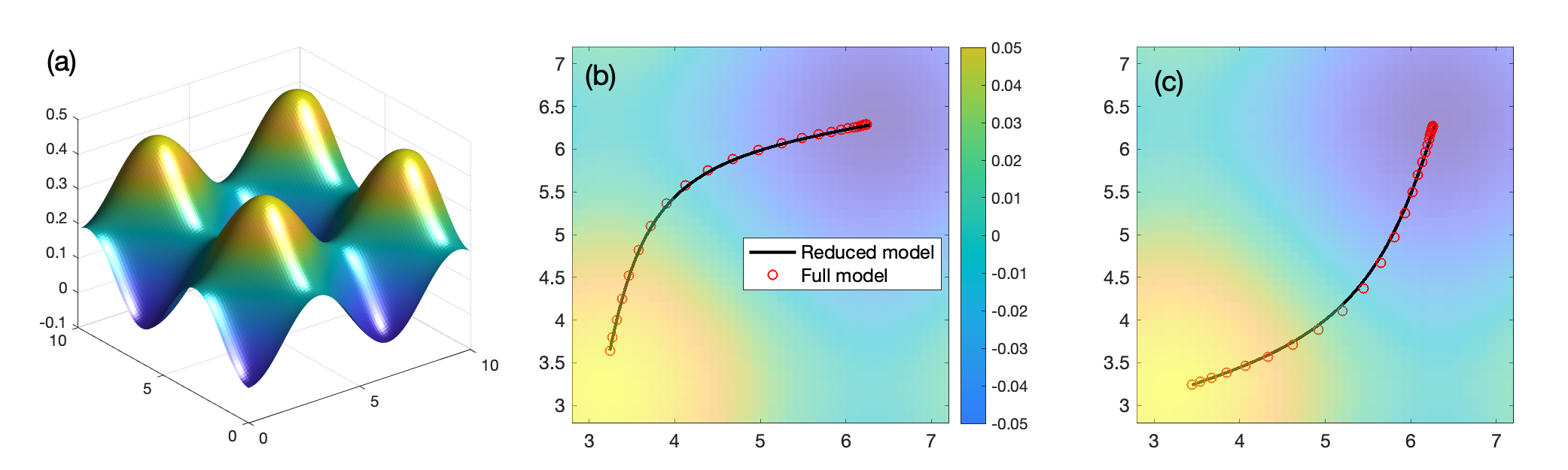}
\includegraphics[width=0.9\textwidth]{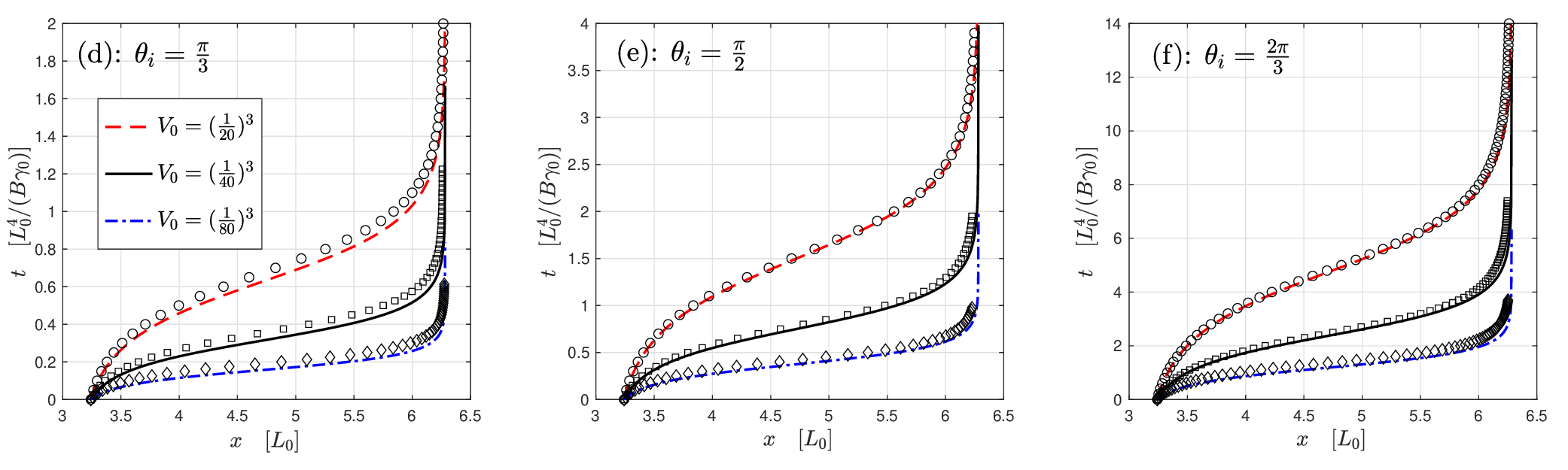}
\caption{(a) An illustration of the egg-carton shape substrate surface; (b) and (c) are the trajectories of the particle starting from different initial positions, where the solid black lines and red circles represent results from the reduced  and full models, respectively, and the shading colors represent the contour of the mean curvature; (d)-(f) show  the $x$-positions of  particles with different initial volumes $V_0$ and Young angles, where the symbols and curves represent  full and reduced model results according to Eq.~\eqref{eq:3ddynamic}. }
\label{fig:6}
\end{figure*}

To numerically confirm the reduced model for particle motion in 3D, Eq.~\eqref{eq:3ddynamic}, we  compare these and the full model, Eq.~\eqref{eqn:fmod}.
Similar to the 2D case, we normalize all lengths by $L_0$ and times by $\frac{L_0^4}{B\gamma_0}$ such that the reduced model can be written in the dimensionless form as
\begin{equation}\label{eq:d3ddynamic}
\begin{cases}
    \displaystyle \frac {{\rm d} \hat{\vec P}}{{\rm d} \hat{t}} = - \frac{C_{3\rd}(\theta_i)}{\sqrt[3]{\hat{V}_0}}\,\hat{\nabla}_\Gamma\hat{\kappa}(\hat{\vec P}),\\
    \hat{\vec P}(0) = \hat{\vec P}_0.
\end{cases}
\end{equation}
We solve for the particle trajectory via the forward Euler method.
For the full model, as described in \ref{app:computfmod}, we employ parametric finite element approximations - see ~\cite[Remark 3.5]{BGNZ23}.
As shown in Fig.~\ref{fig:6}(a), we consider an egg-carton shape  substrate surface
\begin{equation}
z(x,y) = \frac{1}{5}[\sin^2(\frac{x}{2})+\sin^2(\frac{y}{2})].\nn
\end{equation}
We choose $\theta_i=\frac{\pi}{2}$ and place a small volume particle  ($\hat{V}_0=20^{-3}$)  at the point $\bigl(x_0, y_0, z(x_0,y_0)\bigr)$.
In our computation, we start with two different initial positions: $(x_0,y_0)=(\pi+\frac{1}{10}, \pi+\frac{1}{2})$ and $(x_0,y_0)=(\pi+\frac{3}{10}, \pi+\frac{1}{10})$.
The trajectory of the particle in the reduced and full models are compared in Fig.~\ref{fig:6}(b) and Fig.~\ref{fig:6}(c).
Note that the two models show  excellent agreement over the entire trajectories,  confirming our reduced model for the dynamics of  the particle in 3D.

To further assess the reduced model, we next consider the more general cases by varying the volume of the particle and Young angle $\theta_i$.
We compare the time history of the $x$-position of the particle (the initial particle position was $(x_0,y_0)=(\pi+\frac{1}{10}, \pi+\frac{1}{2})$) from the two models by fitting $m(\theta_i)$ in \eqref{eq:3dPhidot}.
The results are shown in Figs.~\ref{fig:6}(d)-(f).
Again,  we observe excellent consistency between the full and reduced models.

\begin{figure*}[!htp]
\centering
\includegraphics[width=0.9\textwidth]{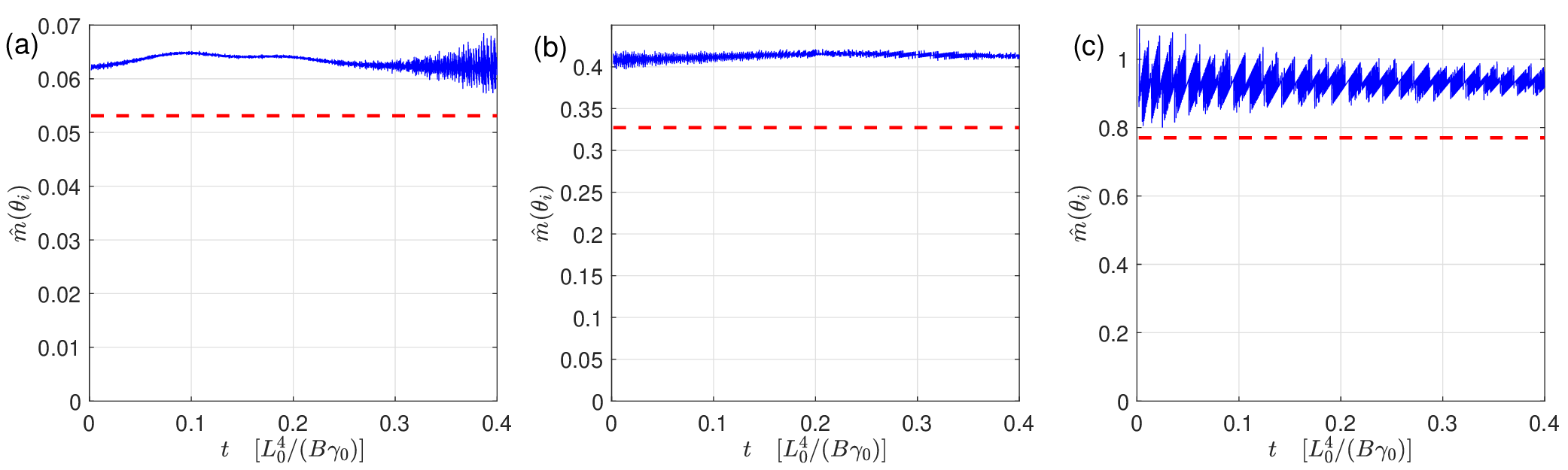}
\caption{The dimensionless function $\hat{m}(\theta_i)$ versus time (blue line) from the  full model  \eqref{eqn:fmod} simulation. The red dashed line is $m(\theta_i)$ from the reduced model \eqref{eqn:cmin}, where   $\hat{V}_0=20^{-3}$ and (a) $\theta_i={\pi}/{3}$, (b) ${\pi}/{2}$, and (c) ${2\pi}/{3}$  are used.  }
\label{fig:8}
\end{figure*}

In our final numerical test, we focus on the verification of  the dissipation function \eqref{eq:3dPhidot} in the reduced model.
$m(\theta_i)$ in \eqref{eq:3dPhidot} can be computed in a minimization framework, as discussed in \ref{app:cmin}.
For the full model,  we introduce an analogue of $m(\theta_i)$ as
\begin{equation}
\hat{m}(\theta_i)= \frac{1}{2\,|\frac{\rd\hat{\vec P}}{\rd \hat{t}}|^2}\left(\frac{\pi\,\eta(\theta_i)}{3\,\hat{V}_0}\right)^\frac{4}{3}\int_{\hat{\mS}(t)}|\hat{\nabla}_s\mathcal{\hat{H}}|^2\,\rd {\hat{S}}.
\end{equation}
If the dissipation law for the migrating particle is adequately approximated by \eqref{eq:3dPhidot}, then $\hat{m}(\theta_i)$ should be similar to the constant $m(\theta_i)$ in time.
We test this for several different Young angles ($V_0=\frac{L_0^3}{80^3}$).
The time history of $\hat{m}(\theta)$ as well as the constant $m(\theta_i)$ are shown in Fig.~\ref{fig:8}.
Indeed, we observe a small oscillation of $\hat{m}(\theta_i)$ about a constant which is  slightly larger than $m(\theta_i)$.
These discrepancies may be associated with the errors in the asymptotic approximations in the reduced model or  numerical  errors in the computation of the full sharp-interface model, see \ref{app:computfmod}.

\begin{figure}
    \centering
    \includegraphics[width=0.5\textwidth]{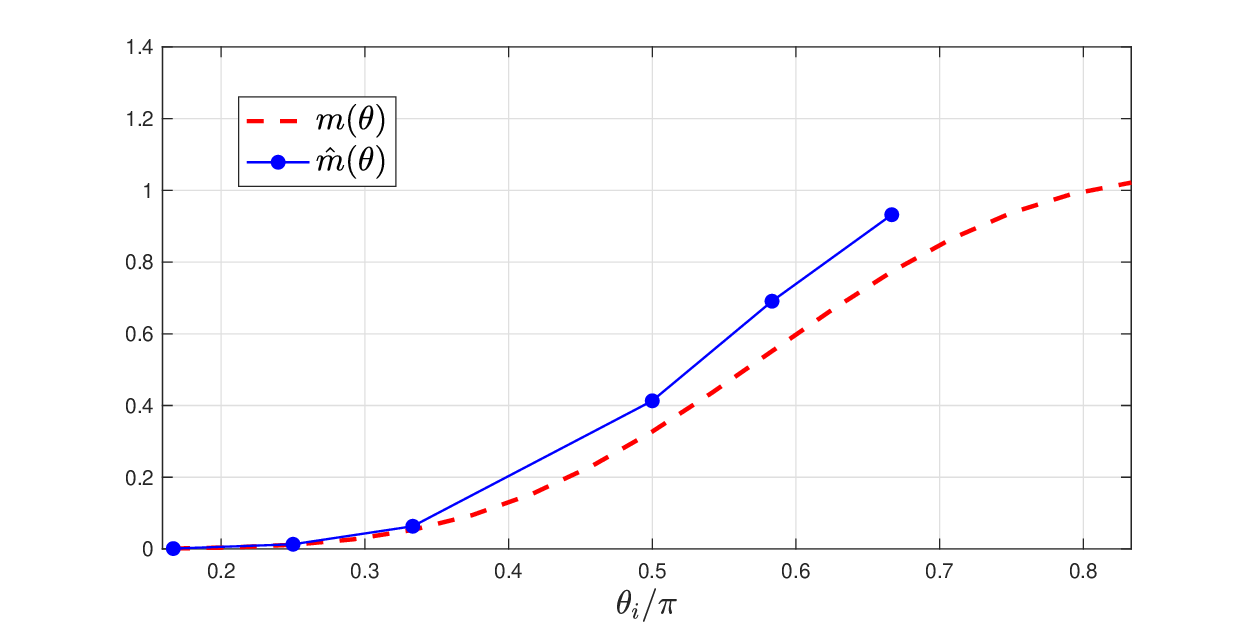}
    \caption{Comparison between $m(\theta_i)$ and $\hat{m}(\theta_i)$ as functions of $\theta_i$.}
    \label{fig:mtheta}
\end{figure}

Finally, we assess the quantitative agreement between $m(\theta_i)$ and $\hat{m}(\theta_i)$ as functions of $\theta_i$ in Fig.~\ref{fig:mtheta}.
Again, the agreement is good  and consistent {\it across two orders of magnitude} from $10^{-2}$ to $1$. Furthermore, $m(\theta_i)$, obtained from the constrained minimization, is everywhere very close to $\hat{m}(\theta_i)$ from the full model computation, as expected since $m(\theta_i)$ is a theoretical value in the minimum dissipation framework, while $\hat{m}(\theta_i)$ is  a numerical value measured in simulation.  The relative error  may be attributed to the following:
(i) In the full model computation, the particle is not small enough compared to
the radius of  curvature  of the substrate, implying that the equilibration of the particle shape is not fast enough compared with the migration along the substrate as assumed in the reduced model;
(ii) The numerical results produced in the full model computation are not sufficiently accurate;
we  observe that when the parameter $\theta_i$ is near $\pi$ (e.g., $\frac{3\pi}{4}$),
a large oscillation in $\hat{m}(\theta_i)$ occurs which may make the full model numerical computation (see \ref{app:computfmod}) unreliable.
This further suggests the utility of the reduced-order model  \eqref{eq:3ddynamic} for such particle migration studies.

\section{Generalizations}\label{SEC:GEN}

\subsection{For chemically inhomogeneous substrates}

We employed our framework of the Onsager principle to the case of deposited thin films on a flat and chemically inhomogeneous substrate. This will drive the particle to transport over the substrate in a similar behaviour of the droplet, see \cite{Malinowski20advances}. We assume that the material parameter $\theta_i$ of the substrate is a function of the position, and is varying pretty slowly. Thus there is a fast time scale for the thin film to form a circular shape and a slow time scale for the migration of the particle.

In the 2D case, using \eqref{eq:2dvol} and \eqref{eq:2denergy} and letting $R\to+\infty$, it is not difficult to obtain that
\begin{equation}
W(P) = 2\gamma_0\,\sqrt{A_0\,\zeta(\theta_i)},\nn 
\end{equation}
of which taking time derivative gives us
\begin{equation}
\dot{W}(P,\dot{P}) = \gamma_0\, \sqrt{A_0}\,\frac{\zeta^\prime(\theta_i)\,\theta_i^\prime(P)\,\dot{P}}{\sqrt{\zeta(\theta_i)}}.
\label{eq:2denergytime}
\end{equation}
We note that the dissipation function for the interface will stay unchanged and can be computed by \eqref{eq:finalPhi} as well. Now combining \eqref{eq:2denergytime} and \eqref{eq:finalPhi} and  applying the Onsager principle yields the following ODE for $P$,
\begin{equation}
\begin{cases}
    \displaystyle \frac {{\rm d} P}{{\rm d} t} = - \frac{1}{2}\frac{B\,\gamma_0}{A_0}\,\frac{\zeta(\theta_i)\,\zeta^\prime(\theta_i)}{C_{2{\rm d}}^0(\theta_i)}\,\theta_i^\prime(P),\\
    P(0) = P_0.
\end{cases}
\end{equation}
This implies that the velocity of the particle is proportional to the gradient of the Young's angle $\theta_i$. Moreover, the parameter $\theta_i$ will also determine how fast the particle move over the substrate.

For the 3D particle on the flat and chemically inhomogeneous substrates, on recalling \eqref{eq:3denergyapp}, we have
\begin{equation}
W(\vec P) = 2\gamma_0\sqrt[3]{9\,V_0^2\,\pi}\,C_0(\theta_i),
\end{equation}
where
\[C_0(\theta_i) = [\eta(\theta_i)]^{-\frac{2}{3}}(1-\cos\theta_i-\frac{1}{2}\cos\theta_i\sin^2\theta_i).\]
Taking time derivative of $W(\vec P)$ then leads to
\begin{equation}
\dot{W}(\vec P, \dot{\vec P}) = 2\gamma_0\,\sqrt[3]{9\,V_0^2\,\pi}\,C_0^\prime(\theta_i)\nabla_\Gamma\theta_i(\vec P)\cdot\dot{\vec P},
\end{equation}
where $\nabla_\Gamma$ is again the gradient along the substrate surface. On recalling the dissipation function \eqref{eq:3dPhidot} in 3D and using the Onsager principle, we obtain the following ODE system for the particle velocity
\begin{equation}
\begin{cases}
    \displaystyle \frac {{\rm d} \vec P}{{\rm d} t} = - \frac{B\gamma_0}{\sqrt[3]{V_0^2}}\,\left(\frac{\pi^5\eta^4(\theta_i)}{9\,m^3(\theta_i)}\right)^\frac{1}{3}C_0^\prime(\theta_i)\,\nabla_\Gamma\theta_i(\vec P),\\
    \vec P(0) = \vec P_0,
\end{cases}
\end{equation}
which implies that the particle moves in a direction that is aligned with the gradient of the parameter $\theta_i$.

\subsection{Discussions on anisotropic surface energies}
We next consider the case when the surface energy of the thin film is anisotropic and modeled by a convex anisotropy function $\gamma(\vec n)$. Again, we assume that the size of the particle is much smaller than the curvature radius of the substrate surface so that there exists a short time scale for the particle to form an anisotropic quasi-static shape and a long time scale for the particle to migrate along the substrate.

On the short time scale, we assume that the particle forms a shape that is no longer circular or spherical but depends on $\gamma(\vec n)$ and also the local substrate topology, i.e., the unit normal $\vec n_w$ to the substrate surface.
Starting from a flat substrate surface with a constant $\vec n_w$, we note that the zeroth-order particle shape 
$\mathcal{S}_\gamma$ can be constructed via the Winterbottom construction \cite{Winterbottom67} or the well-known Cahn-Hoffman vector formulation \cite{Hoffman72}.
As the substrate surface becomes gently curved with a nonzero $\nabla_\Gamma\vec n_w$, 
the particle shape and the free energy $W(\vec P)$ will be slightly modified by $\nabla_\Gamma\vec n_w$. 
To the first order, we consider the linear dependence of $W(\vec P)$ on $\nabla_\Gamma\vec n_w$ by noting that 
the orientation of $\mathcal{S}_\gamma$ relative to the principal axes of $\nabla_\Gamma\vec n_w$ is involved. 
This is totally different from the isotropic case, where the particle shape possesses a rotational symmetry so that the linear dependence of $W(\vec P)$ on $\nabla_\Gamma\vec n_w$ is
through $\mathrm{Tr}[\nabla_\Gamma\vec n_w]$, i.e., the mean curvature of the substrate surface, see \eqref{eq:3denergyapp}.

To compute the dissipation function $\Phi(\dot{\vec P})$,  we can employ a technique similar to the isotropic case in the framework of constrained minimum dissipation. This leads to the minimization problem \eqref{eqn:cmin}, 
where the anisotropic effects are manifested in the integration over the anisotropic interface $\mathcal{S}_\gamma$ 
and the constraint which involves the unit normal $\vec n$ of $\mathcal{S}_\gamma$.

In summary, our variational reduced-order modeling approach can also be employed to deal with solid thin films with anisotropic surface energies on topologically patterned substrates, although 
the particle migration will exhibit much more complicated trajectories.

\section{Conclusions}\label{SEC:CON}

We studied the dynamics of a small particle on  curved substrates driven by surface/interface energies and controlled by surface diffusion  with a moving contact line.
We observed two distinct time scales for the dynamical behaviour of the particle.
Small particles evolves towards their equilibrium shape (circular or spherical section in 2D or
 3D) on a fast time scale.
The small particle move  along the substrate in order to adapt itself to the substrate topography on a slow time scale.
We derived a reduced-order model to describe this particle migration using the Onsager variational principle.
The particle moves in a direction in which the curvature of the substrate decays most quickly.
The particle velocity is proportional to a material constant $B=\frac{D_s\nu\Omega_0^2}{k_b\,T}$, the surface energy density $\gamma_0$, and inversely proportional to the size of the particle.
The reduced model was numerically confirmed by comparing with  numerical results obtained by numerical solution of the full dynamical model.

The main overall conclusion is that the trajectory of  small particle on a substrate is solely determined by the substrate topography, while only the rate of motion is controlled by material properties. 
This may provide some insights into the understanding of the templated dewetting on  patterned substrate.

\section*{Acknowledgements}
This work was partially supported by the National Natural Science Foundation of China Nos. 12271414 and 11871384 (W.J.) and 
No. 12371395 (Y.W.), the NSF Division of Materials Research through Award 1507013 (D.J.S.), 
Hong Kong RGC grants CRF No. C1006-20WF and GRF No. 16306121 (T.Q.), 
and the Ministry of Education of Singapore under its AcRF Tier 2 
funding MOE-T2EP20122-0002 (A-8000962-00-00) (W.B.).

\appendix

\section{Computational Method for the Full Model \eqref{eqn:fmod}}\label{app:computfmod}
The computation of the full model \eqref{eqn:dfmod} is based on the following formulation  \cite{Barrett20}:
\begin{subequations}\label{eqn:dsf}
\begin{align}
\hat{\vec X}\cdot\vec n = \hat{\Delta}_s\mathcal{\hat{H}},\\
\mathcal{\hat{H}}\vec n = -\hat{\Delta}_s\hat{\vec X}.
\end{align}
\end{subequations}
A weak formulation is introduced for \eqref{eqn:dsf},  where the contact angle condition \eqref{eq:dbd2} and the zero-flux condition \eqref{eq:dbd3} are implemented via the variational formulation (see \cite{BGNZ23} for details).
To  ensure satisfaction of the attachment condition \eqref{eq:dbd1}, the velocity of the contact line is forced to be tangential to the substrate surface.
We  employ a piecewise linear element method in space and a backward Euler discretization in time to obtain a parametric approximation, where the interface surface is approximated as a polyhedron.
This  discretization implies that  contact points may not exactly lie on the curved substrate at the following time step.
Therefore, an orthogonal projection of these points onto the substrate surface is required; this may incur some additional numerical errors beyond that associated with the discretization of the geometric equation.

\section{The Dissipation Function}\label{app:cmin}
We  introduce a dimensionless flux on a spherical section of the surface
\begin{equation}
    \tilde{\vec j} = \mF(\theta,\phi)\,\vec e_\theta +\mG(\theta,\phi)\vec e_\phi,
\end{equation}
where $\vec e_\theta$ and $\vec e_\phi$ represent the unit vector in the axial and azimuthal directions, respectively.

To calculate the dissipation function in 3D, we consider the constrained minimization problem \eqref{eqn:cmin} which is rewritten in the spherical coordinate and in terms of $\tilde{\vec j}$ as
\begin{subequations}\label{eqn:sphcmin}
\begin{align}
&\min_{\tilde{\vec j}}\frac{1}{2}\frac{k_b\,T\,r^4}{D_s\,\nu\,\Omega_0^2}\int_0^{2\pi}\int_0^{\theta_i}(\mF^2 + \mG^2)\sin\theta\rd\theta\rd\phi,\\
&\mbox{with}\;\frac{\partial(\mF\,\sin\theta)}{\partial\theta} +\frac{\partial \mG}{\partial\phi} = -(\frac{\dot{\vec P}}{|\dot{\vec P}|}\cdot\vec n^\phi)\sin^2\theta,\\
&\mbox{and}\quad\mF(\theta_i,\phi) = 0,
\end{align}
\end{subequations}
for $\theta\in[0,\theta_i]$ and $\phi\in[0,2\pi]$, where $\vec n^\phi = (\cos\phi, \sin\phi, 0)^T$.

We then discretize the problem in $(\theta,\phi)\in[0,\theta_i]\times[0,2\pi]$.
We introduce the following notations:
\begin{subequations}
\begin{align}
h_\theta = \frac{\theta_i}{N}, \quad \theta_j = j\,h_\theta\quad\mbox{for}\quad 0\leq j\leq N, \\
\qquad h_\phi = \frac{2\pi}{M},\quad \phi_k = k\,h_\phi\quad\mbox{for}\quad 0\leq k\leq M,
\end{align}
\end{subequations}
and $\mF_{jk}\approx\mF(\theta_j,\phi_k)$, $\mG_{jk}\approx\mG(\theta_j,\phi_k)$ and  rewrite \eqref{eqn:sphcmin} in the discrete form.
In particular, the objective function, up to a multiplicative constant, is approximated as
\begin{align}\label{eq:dobj}
m(\theta_i)&=\frac{1}{2}\int_0^{2\pi}\int_0^{\theta_i}(\mF^2+\mG^2)\,\sin\theta\rd\theta\rd\phi\nn\\ &\approx\frac{1}{2}\sum_{j=1}^N\sum_{k=1}^M(\bar{\mF}_{jk} + \bar{\mG}_{jk})\frac{2\pi \theta_i}{NM},
\end{align}
where
\begin{align}
\bar{\mF}_{jk}=\frac{1}{4}\sum_{l_1=0,1}\sum_{l_2=0,1} \mF_{(j-l_1)(k-l_2)}^2\sin\theta_{j-l_1},
\end{align}
and $\bar{\mG}_{jk}$ is defined similarly.

We  introduce the finite difference discretization operators as follows:
\begin{subequations}
\begin{align}
\delta_\theta\mA_{jk}&=\frac{\mA_{j+1,k}-\mA_{j-1,k}}{2h_\theta},\quad \delta_\phi\mA_{jk}=\frac{\mA_{j,k+1}-\mA_{j,k-1}}{2h_\phi},\nn\\
\delta_\theta^+\mA_{jk}&=\frac{\mA_{j+1,k}-\mA_{j,k}}{h_\theta},\quad \delta_\theta^-\mA_{jk}=\frac{\mA_{j,k}-\mA_{j-1,k}}{h_\theta},\nn
\end{align}
\end{subequations}
and observe that  $\mF_{0,k}$ and $\mG_{0,k}$ make no contributions to the objective function (because $\sin0=0$).
The constraints are discretized naturally as
\begin{itemize}
\item for $j=1$, $1\leq k\leq M$,
\begin{subequations}\label{eq:dcst}
\begin{align}
\delta_\theta^+(\mF\sin\theta)_{1k}+\delta_\phi\mG_{1k}=-\left[\frac{\dot{
\vec P}}{|\dot{\vec P}|}\cdot\vec n^{\phi}_{1k}\right]\sin^2\theta_{1k},
\end{align}
\item for $1< j < N$ and $1\leq k\leq M$,
\begin{align}
\delta_\theta(\mF\sin\theta)_{jk}+\delta_\phi\mG_{jk}=-\left[\frac{\dot{
\vec P}}{|\dot{\vec P}|}\cdot\vec n^{\phi}_{jk}\right]\sin^2\theta_{jk},
\end{align}
\item  for $j=N$, $1\leq k\leq M$,
\begin{align}
\mF_{N,k}=0.
\end{align}
\end{subequations}
\end{itemize}

This is the discrete minimization problem for the objective function \eqref{eq:dobj} with the constraints \eqref{eq:dcst}; it is a quadratic minimisation problem with linear constraints for $\{\mF_{jk}\}$ and $\{\mG_{jk}\}$ for $j=1,\ldots, N$ and $k=1,\ldots, M$.
Thus it can be solved directly as a linear system corresponding to the Karush-Kuhn-Tucker (KKT) condition.

\end{document}